\documentclass{aa}
\usepackage{times,graphicx,amsmath,amssymb,mathtools,ulem,widetext,hyperref}

\usepackage{xcolor,soul}

\DeclareMathAlphabet\mathbfcal{OMS}{cmsy}{b}{n}
\begin{document}

\title{Depolarization and polarization transfer rates for the C$_2$ $(X ^1\Sigma^+_g, a ^3\Pi_u)$ + H$(^2S_{1/2})$ collisions in the solar photosphere}
\author{ S. Qutub$^{1}$,   Y.N. Kalugina$^{2,3}$, M. Derouich$^{1}$}

\institute{ 
$^{1}$ Astronomy and Space Science Department, Faculty of Science, King Abdulaziz University, P.O. Box 80203, Jeddah 21589, Saudi Arabia   \\
$^{2}$ 
Department of Optics and Spectroscopy, Tomsk State University, 36 Lenin av., Tomsk 634050, Russia \\
$^{3}$ Institute of Spectroscopy, Russian Academy of Sciences, Fizicheskaya St. 5, 108840 Troitsk, Moscow, Russia  
}

\titlerunning{Depolarization and polarization transfer rates for the C$_2$}
\authorrunning{Qutub et al.}

\date{XXXX / XXXX }

\abstract
  {This paper is a continuation of a series of studies investigating collisional depolarization  of solar molecular lines like those of MgH, CN  and C$_2$. It is focused on the case of the solar molecule C$_2$  which exhibits striking scattering polarization profiles although   its intensity profiles are  inconspicuous
and barely visible.  In fact, interpretation of the C$_2$  polarization 
in terms of magnetic fields is  incomplete due to  the almost complete lack of collisional data.
}
%
%
{   This work aims at accurately computing the collisional depolarization and polarization transfer rates for the C$_2$~$(X ^1\Sigma^+_g, a ^3\Pi_u)$  by isotropic collisions with hydrogen atoms H~$(^2S_{1/2})$. We also investigate   the solar implications of our findings.
}
{We utilize the MOLPRO package to obtain potential energy surfaces (PESs) for the electronic states $X ^1\Sigma^+_g$ and $a ^3\Pi_u$  of C$_2$,  
and the MOLSCAT code to study the quantum dynamics of the C$_2$~$(X ^1\Sigma^+_g, a ^3\Pi_u)$ + H$(^2S_{1/2})$ systems. We use the  tensorial irreducible basis to express
the resulting
collisional cross-sections and rates.  
Furthermore, sophisticated genetic programming techniques are employed to determine analytical expressions for the temperature and total molecular angular momentum dependence of these collisional rates.}
{
{ We obtain quantum depolarization and polarization transfer rates for the C$_2$ $(X ^1\Sigma^+_g, a ^3\Pi_u)$ + H$(^2S_{1/2})$ collisions   in the temperature range  $T \!=\!$~2,000~--~15,000~K. We also determine analytical expressions giving these rates as functions of the temperature and total molecular angular momentum.
 In addition, we    show that isotropic collisions with neutral hydrogen can only partially depolarize the lower state of C$_2$ lines, rather than completely. This highlights the limitations of the approximation of neglecting lower-level polarization while modeling the polarization of C$_2$ lines. }  
 }
{Isotropic collisions with neutral hydrogen atoms are fundamental ingredient   for understanding C$_2$ polarization.     }


\offprints{M. Derouich, \email{DerouichMoncef@gmail.com}}

\maketitle

\section{Introduction} \label{zero-magnetic field case}
Highly sensitive spectro-polarimetric telescopes have opened new observation windows on the lines of MgH, CN, and C$_2$ molecules with unprecedented spatial and spectral resolutions
 (e.g.  Stenflo, 1994; Gandorfer 2000; Berdyugina, Stenflo, \& Gandorfer  2002;  Faurobert  \& Arnaud  2003; Berdyugina \& Fluri 2004; Asensio Ramos  \& Trujillo Bueno 2005;  Trujillo Bueno et al. 2006;    Mili\'c \& Faurobert   2012; Wiegelmann et al. 2014;  W\"oger et al. 2021). 
Furthermore, Hanle effect
 on molecular polarized solar lines provides a good opportunity to determine spatially unresolved magnetic fields given the diverse magnetic sensitivities of molecular lines observed within  narrow spectral regions. 
In fact, observation and interpretation of  
molecular C$_2$ lines of the   Swan system  ($d ^3\Pi_u$ -- $a ^3\Pi_u$) around 5141  \AA \;  
constitute an interesting 
tool 
to infer the magnetic field strength   (e.g. Berdyugina \& Fluri 2004;  Mili\'c  \&  Faurobert 2012).  
Nevertheless, 
some discrepancies have been found in the results regarding magnetic field strengths (e.g. Asensio Ramos  \& Trujillo Bueno 2005; Derouich et al. 2006; Kleint et al. 2010).  
To eliminate
a primary cause of these discrepancies, the  effect of collisions in the formation of molecular lines  should be taken into account. 
In fact, 
the main difficulty facing all Hanle diagnostics of the molecular lines is  that  collisional rates are poorly known.  
 Particularly, collisions  with hydrogen atom are of great importance due to 
its high density in the photosphere where Hanle effect is in action.
Ignoring collisions has a significant consequence on the precise determination of magnetic fields, as collisions compete with the Hanle depolarizing effect of the turbulent photospheric magnetic fields.   To contribute in addressing this difficulty,
 Qutub et al. (2020,  2021)    calculated, 
 for the first time,
the collisional depolarization and polarization transfer rates of the ground states of the MgH and CN molecules by collisions with hydrogen atoms.  
In a continuation of this effort, we aim to compute the depolarization and transfer of polarization rates for the two lowest energy electronic states of C$_2$ due to collisions with hydrogen atoms.

As C$_2$ is a homonuclear molecule, 
transitions between rotational  levels with different parities cannot occur (see e.g.~Flower 1990; Derouich 2006). 
This restriction arises from the symmetry of the interaction potential. 
However, this does not imply that C$_2$ molecule is immune to collisions.
As it will be demonstrated through this paper, 
collisional transitions between rotational levels with same parity can affect the polarization of the C$_2$ electronic levels, and, therefore affect the polarization of the C$_2$ solar lines (e.g. Kleint et al. 2010). 
In this regard,  we  calculate the depolarization 
and  polarization transfer rates of the
solar C$_2$ molecules in their ground and first excited states $X ^1\Sigma^+_g$ and  $a ^3\Pi_u$
due to collisions with
  H atoms in their ground state $^2S_{1/2}$.

The first step of this calculation 
is the determination of potential energy surfaces (PESs) for H+C$_2$ interactions. 
All 
the PESs are obtained using the MOLPRO package (e.g. Werner et al. 2010).
The second step is
 the 
determination of the dynamics of collisions by solving the corresponding
 Schr\"odinger equations. 
The dynamics calculations are made possible thanks to the MOLSCAT code (e.g. Hutson \& Green 1994). 
Then, 
the depolarization and polarization transfer cross-sections are computed within the tensorial basis $T^k_q$ where $k$ is the tensorial order and $q$ quantifies the coherence within the molecular level. 
Note that 
these cross-sections are $q$-independent since the collisions are isotropic. 
We 
adopt the infinite order sudden (IOS) approximation to calculate the cross-sections for kinetic energies ranging from 50 to 40,000 cm$^{-1}$ allowing
the calculation of depolarization and polarization transfer rates  for temperatures ranging from $T\!=\!$ 2,000 to 15,000 K. 
Finally,
  genetic programming methods are applied to infer useful analytical expressions of the obtained rates. 
In addition, the
expected solar implications of our results are  briefly discussed.
Our cross sections are available online for future use by the community.

\section{Potential Energy Surfaces} \label{PESs}
We consider the   ground $X ^1\Sigma^+_g$ and the first excited $a ^3\Pi_u$  electronic states of C$_2$ molecule which are
close in energy  with spacing between the two states of about only 700 cm$^{-1}$ (i.e. $\sim$ 0.087 eV) (see e.g. Martin 1992). 
When
C$_2$ ($X ^1\Sigma^+_g$)  interacts with the hydrogen atom H in its $^2S$ ground electronic state, the resultant system can exist in one electronic state  $1  \; ^2A'$. 
Furthermore,
 $2 \; ^2A'$ and $2  \; ^2A''$ represent the  two states which  result from  the interaction between C$_2$($a ^3\Pi_u$) and H ($^2S$). We adopt the coordinate system of Jacobi ($R$, $r_{C_2}$, $\theta$)
for the calculation of the PESs. 
The intermolecular vector \textbf{$R$}
connects the center of mass of C$_2$ molecule and the hydrogen atom. The angle $\theta$ defines the rotation of the hydrogen atom around the C$_2$ molecule. 
In the present work,
the C$_2$ molecule is assumed to be rigid rotor 
with C-C distance frozen at its equilibrium value $r_{C_2}\!=\! 2.348$~$a_0$ (Huber \&  Herzberg 1979). 
This is justified in the solar physical conditions where the rates for vibrational excitations due to collisions are much smaller than those for pure rotational excitations.

Ab initio calculations of the PESs for the  
electronic states of C$_2$-H system, described above,   were carried out using 
the multi-reference configuration interaction
wave functions including Davidson correction (MRCI+Q) 
 (see   Langhoff \& Davidson 1974; Davidson \& Silver 1977;  Werner \&
Meyer 1981;  Wener  \& Knowles 1988). 
The
computations were performed using the MOLPRO 2010 package (e.g. Werner et al. 2010).

\begin{figure}
\includegraphics[width=9.5cm]{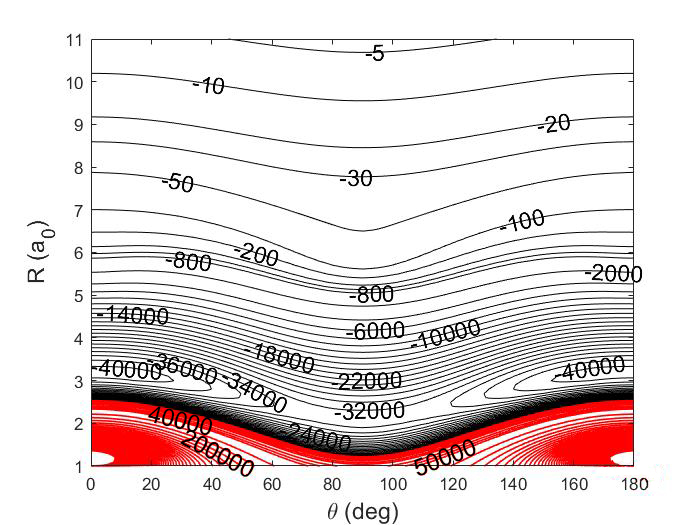}   
\caption{Contour   plot  of the PES  of the electronic state   $1  \; ^2A'$ as a function of $R$ and $\theta$. Energy is in  cm$^{-1}$.}
\label{fig:pot3d_s_1}
\end{figure}

\begin{figure}
\includegraphics[width=9.5cm]{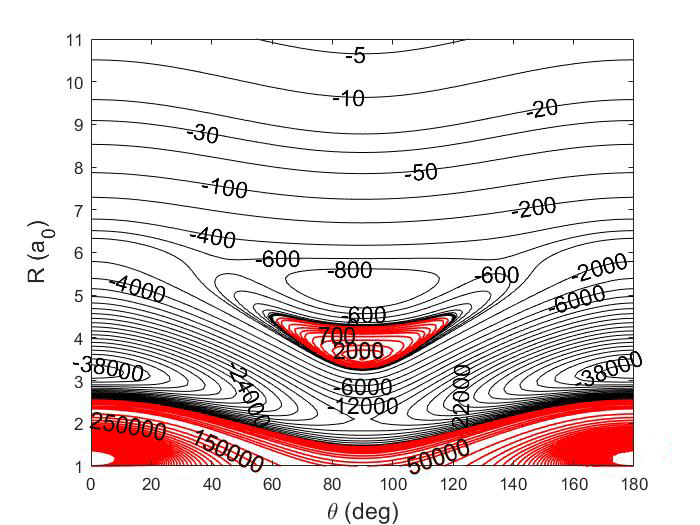}   
\includegraphics[width=9.5cm]{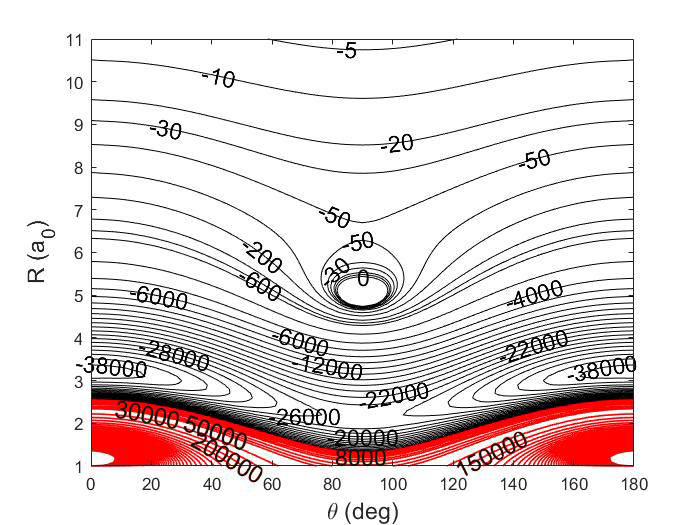} 
\caption{Contour   plots of the PESs of the electronic states   $2 \; ^2A'$ (upper panel) and $2  \; ^2A''$ (lower panel) as functions of $R$ and $\theta$. Energy is in  cm$^{-1}$.}
\label{fig:pot3d_s_2}
\end{figure}

For the electronic states under consideration, the two dimensional PESs were generated for an  angle  $\theta$ ranging from 0 to $90^\circ$ by using a variable step in order to well cover the
 behavior of the PESs and by varying 
 the $R$ values  from
 $a_0$ to 50 $a_0$.
  We used 84 values of $R$ 
 ($a_0 \!\le\! R \!\le\! 50~a_0$) 
 and 51 values of $\theta$ 
 ($0^\circ \!\le\! \theta \!\le\! 90^\circ$) 
  implying that the total number of the generated ab initio points is 4284 for each potential surface  V($R$, $\theta$). 
Note that,
since C$_2$ is a homonuclear molecule, the interaction potential V is  invariant under exchange of the two carbon atoms,
 i.e.,  
 V($R$,~$\theta$)~=~V($R$,~$180^\circ \!- \theta$).  
The
PES of  the electronic state  $1  \; ^2A'$, resulting from C$_2$ ($X ^1\Sigma^+_g$)  and  H ($^2S$) interaction,
 is  
shown in Figure \ref{fig:pot3d_s_1}.
Furthermore, 
PESs of  the  $2 \; ^2A'$ and $2  \; ^2A''$ states represented in the  Figure \ref{fig:pot3d_s_2}  result from  the interaction between C$_2$($a ^3\Pi_u$) and H ($^2S$).

The PES corresponding to the state $1 \; ^2A'$ reaches its minimum at $R \!\sim\! 3.2~a_0$ and $\theta \!\sim\! 0^\circ$, with an energy of 
 $E \!\sim\!  \text{-41,000}$~cm$^{-1}$.  The PESs corresponding to the state $2 \; ^2A'$ and  $2 \; ^2A''$ states  have a similar minimal energy 
$E \!\sim\! \text{-39,000}$~cm$^{-1}$ 
with
 $\theta \!\sim\! 0^\circ$ and $R \!\sim\! 3.2~a_0$.

\section{Collisional problem}
In the context of the close coupling (CC) scheme, generating comprehensive results for depolarization and polarization transfer rates of the C$_2$ molecule is very difficult 
given the
 numerous   rotational levels,
the spin characteristics, and
the 
large number of possible  $k$-values for each  rotational level. To overcome this difficulty we adopt the
IOS
approximation, which is appropriate for solar temperatures,
to treat the collision problem and provide a comprehensive data for all collisional rates.
As we are interested in the solar context, where the temperature and the kinetic energies of collisions are sufficiently high, one can expect that some simplification regarding the coupling effects should be invoked in order to obtain results with acceptable accuracy in a reasonable  calculation time.

We adopt  a formalism developed  by Corey \& Alexander (1985) which is 
based on the decoupling of the rotational-orbital motion from the atomic and molecular spin angular momenta  (see also Corey \& McCourt 1983).
This decoupling scheme offers an advantage in our attempt to express the tensorial cross-sections in terms of the generalized
IOS cross-sections especially for the case of  C$_2$($a ^3\Pi_u$) state where the orbital angular momentum and the spin of the molecule are non-zero.  
It is worth mentioning that, 
in a collision between an open-shell molecule (like  C$_2$ $a ^3\Pi_u$) and an open-shell target  (like H  $^2S$) the PESs is dependent of the total spin of the composite atom-molecule 
system and that dependence is taken into account in the  PESs calculations. However, we neglect the effect of the spin of the hydrogen in the dynamics of collisions and we only take into account the molecular spin which is a further approximation necessary for expressing  the tensorial cross-sections  factorized into products of terms involving the IOS cross-sections. In these conditions, one has (see Equations 13a,b,c of Corey \& Alexander  1985):
\begin{eqnarray}  \label{eq_coupling}
 { \bf N} &=& { \bf \mathbfcal{R}+L}   \\
 {\bf j} &=& { \bf N+S_d}    \nonumber
\end{eqnarray}
where  $\mathbfcal{R}$ is the rotational angular momentum of the  diatomic C$_2$ molecule
 and
{\bf S$_d$}   its spin;
 $S_d \!=\! 0$ for   $X ^1\Sigma^+_g$
and
 $S_d \!=\!1 $ for $a ^3\Pi_u$.
{\bf L} is the electronic orbital  momentum of the molecular state;
$L \!=\! 0$ for   $X ^1\Sigma^+_g$
 and
$L \!=\! 1$ for $a ^3\Pi_u$. 
Note that ${ \bf N}$ is the total rotational-orbital angular momentum  and ${\bf j} $ is the total momentum of the molecule taking into account the spin.

In the framework of the coupling scheme given in Equation~(\ref{eq_coupling})
 and 
by following a methodology similar to that  explained in different works concerned with molecule-atom collisions and by including the
 IOS
approximation 
(e.g. Pack  1972, 1974; Alexander \& Davis 1983;  Alexander \& Dagdigian 1983; Corey \& Alexander 1985, 1986; Corey \& Smith 1985; Werner, et al. 1989; Follmeg et al. 1990; Green 1994; Dagdigian  \& Alexander 2009 a,b,c;  Paterson et al. 2009; McGurk et al. 2012),
one can show that:
\begin{eqnarray}\label{eq_sigma}
\sigma^k_{IOS}(el, j \to j', E) \!=\! \sum_{\scriptstyle K} 
(-1)^{k+j+j'+K}  (2\mathcal{R}+1)(2\mathcal{R}'+1) \nonumber \\
(2j'+1)(2j+1) (2N+1)(2N'+1)  \nonumber \\
\hspace{-1cm}  \left\{ \begin{array}{ccc} 
j  & j' & K   \\
 j'  & j &k
\end{array}
\right\}    
    \left\{ \begin{array}{ccc} 
 \mathcal{R} & \mathcal{R}' & K  \\
N' & N &L
\end{array}
\right\}^2    \quad \quad \quad  \\
 \left\{ \begin{array}{ccc} 
N  & N' & K  \\
j' & j &S_d
\end{array}
\right\}^2   
 \left( \begin{array}{ccc} 
  \mathcal{R}'& \mathcal{R}  & K \\
 0 &  0 &  0 
\end{array}
\right)^2  \sigma(el, 0  \to     K, E)  \quad \nonumber  \, ,
\end{eqnarray}  
where $E$ is the kinetic energy  and $\sigma_{IOS}^k(el, j \to j', E) $  are the IOS polarization transfer cross sections from the level $(\mathcal{R}Nj)$ to $(\mathcal{R}'N'j')$ within the electronic state $el$ and $ \sigma(el, 0  \to     K, E)$  are the generalized  IOS  cross-sections. 
From
 the general formula of Equation~(\ref{eq_sigma}),
one can recover the limiting case  where   $S_d\!=\!L\!=\!0$   (see e.g. Derouich 2006; Lique et al. 2007).

The depolarization cross-section of the level $(\mathcal{R}Nj)$ is given by:
\begin{equation}\label{eq_5}
\sigma^k_{IOS}(el,j, E)=\sigma^0_{IOS}(el,j \to j, E) -\sigma^k_{IOS}(el,j \to j, E) \, , 
\end{equation}
where $\sigma^k_{IOS}(el,j \to j, E)$ is obtained from Equation~(\ref{eq_sigma}) with $j=j'$, $\mathcal{R}=\mathcal{R}'$, and $N=N'$.

For the resolution of the collision dynamics of the problem at hand under the
IOS approximation,
the PESs were introduced into the MOLSCAT code
(e.g. Hutson \& Green  1994).
As a result,
we obtain  the generalized IOS cross-sections $ \sigma(el, 0  \to     K, E)$ for  energies 50~$\!\le\! E~(\text{cm}^{-1}) \!\le\!$~40,000
 and
  0~$\!\le\! K  \!\le\!$~158 ($K$ is even).
The   data  giving
$ \sigma(el, 0  \to     K, E)$ for all $E$
and 
$K$ values and for $el=a ^3\Pi_u$ and $el=X ^1\Sigma^+_g$ are made accessible online\footnote{Our IOS data are provided to enable reproduction of the IOS rates.}. 
For 
each PES 
we obtain an  IOS cross-section $ \sigma(el, 0  \to     K, E)$, so that we obtain  $ \sigma(1  \; ^2A', 0  \to     K, E)$ for the PES corresponding to the $1  \; ^2A'$  state resulting from C$_2$ ($X ^1\Sigma^+_g$)  and  H ($^2S$) interaction, i.e.,
\begin{equation}\label{eq_5a}
\sigma(el=X ^1\Sigma^+_g, 0  \to   K, E)  =   \sigma(1  \; ^2A', 0  \to     K, E) \, ,
\end{equation}
while   $ \sigma(2 \; ^2A', 0  \to     K, E)$  and  $ \sigma(2  \; ^2A'', 0  \to     K, E)$ are obtained by solving the collision dynamics  after introducing the PESs of  the  $2 \; ^2A'$ and $2  \; ^2A''$ states, respectively.  The     $2 \; ^2A'$ and $2  \; ^2A''$ result from  the interaction between C$_2$($a ^3\Pi_u$) and H ($^2S$) and have the same spin. Thus, in order to obtain the cross-section corresponding to the C$_2$($a ^3\Pi_u$) + H ($^2S$)  one has:
\begin{eqnarray}\label{eq_5b}
\sigma(el=a ^3\Pi_u, 0  \to     K, E)  \quad  \quad  \quad  \quad  \quad  \quad  \quad  \quad  \nonumber \\
   =   \frac{\sigma(2 \; ^2A', 0  \to     K, E)   +   \sigma(2  \; ^2A'', 0  \to     K, E)} {2}  \, .
\end{eqnarray}
The $\sigma_{IOS}^k(el, j \to j', E) $   are then obtained by applying Equation~(\ref{eq_sigma}).  
The  
depolarization rates 
\begin{equation} \label{depolarization}
D^k(el,j,T)  =    D^0(el,j \to j)- D^k(el,j \to j)
\end{equation}
of the  level $(\mathcal{R}Nj)$  due to elastic collisions 
and   the polarization transfer rates $ D^k(el,j \to j', T)$  between the levels  $(\mathcal{R}Nj)$ and $(\mathcal{R}'N'j')$  due to inelastic collisions 
 are obtained by 
 thermally averaging the respective IOS cross-sections:
\begin{eqnarray} \label{gammak}
D^k(el,j \to j', T)  =  n_H \langle \sigma_{IOS}^k(el, j \to j', E) \ v \rangle \\
= n_H \!\left( \frac{8 k_B T}{\pi \mu} \right)^{\!\!\frac{1}{2}} 
\!\!\int_{0}^{\infty} \!\!
\sigma^k(el,j \to j',\epsilon) 
\,  e^{-\epsilon} \epsilon \, d \epsilon \nonumber  
\end{eqnarray} 
for temperatures in the range 2,000~--15,000~K. Here   $n_H$ is the density of the incident hydrogen atoms,  $k_B$ is the Boltzmann constant,  and $\epsilon =  E/k_B T$.

In the case of homonuclear molecules like C$_2$, collisional transitions between levels with even and
odd $j$-values (or $\mathcal{R}$-values in the considered decoupling scheme) cannot occur (see e.g.~Flower 1990; Derouich 2006). 
This fact is 
related to the symmetry of the interaction potential where V($R$, $\theta$)~=~V($R$, $180^\circ \!-  \theta$) (see Section \ref{PESs}). But  obviously,  this does not mean that the C$_2$ molecule is immune to collisions. Collisional transitions
with  $\Delta j$ (or  $\Delta \mathcal{R}$ in our case) even are allowed which constitute a possibility of a collisional contribution to the statistical equilibrium equations (SEE) given by:
\begin{eqnarray} \label{eq_ch3_17}
\big(\frac{d \; ^{j}\rho_q^{k}}{dt})_{coll} & = & - [D^k(j, T)  \nonumber \\ &+& \sum_{j' \ne j}  \sqrt{\frac{2j'+1}{2j+1}} D^0 (j \to  j', T)] \; ^{j}\rho_q^k \nonumber \\
&& + \sum_{j' \ne j}  
D^k(j' \to  j, T) \;  ^{j'}\rho_q^k,  
\end{eqnarray}
where  $^{j}\rho_q^k$ are the density matrix elements expressed in the tensorial basis
which permit a description of the internal states of the C$_2$ molecule       
(e.g. Sahal-Br\'echot 1977; Landi Degl'Innocenti  \& Landolfi 2004). 
The importance of the collisional effects is mainly associated with the value  of $n_H$ and, to a 
lesser extent, with $T$.
In the case of isotropic collisions,
 transfer of polarization rates obey the detailed balance relation:
 \begin{equation}  \label{eq_balance}
D^k(j  \to  j', T) =   \frac{2j'+1}{2j+1}  \exp \left(\frac{E_{j}-E_{j'}}{k_BT}\right)   D^k(j'  \to  j, T) 
\end{equation}
where $E_{j}$ is the energy of the level ($j$).

\section{Results} 

We have determined polarization transfer rates $ D^k(j \to j',T)$ and depolarization rates $ D^k(j,T)$ 
associated to 
rotational levels within C$_2$ electronic states  $X \; ^1\Sigma^+_g$   and   $a \; ^3\Pi_u$.   When possible, genetic programming (GP) fitting techniques were employed  to express these rates   as two-variable functions, with the variables being $j$ which varies from 0 to 60 and $T$ which goes from  2,000 to 15,000~K. The GP expression is:
\begin{equation}  \label{eq_GP}
D^k_{GP}  =  n_{H} \!\times\! 10^{-10} ~~ \frac{\sum_i a_i^k j^{\alpha_i^k} T^{\beta_i^k}}{\sum_i b_i^k j^{\gamma_i^k} T^{\delta_i^k}},
\end{equation}
where the GP coefficients   $a_i^k$,   $\alpha_i^k$,  $\beta_i^k$,  $b_i^k$,   $\gamma_i^k$, and  $\delta_i^k$ are provided in Tables \ref{tab:sigma_depolarization}--\ref{tab:Pi_transfer} of the Appendix A.
As representative examples, 
we provide  $ D^k(j \to j',T)  $ and   $ D^k(j,T)  $ for 6 cases. However,  
Equation~\ref{eq_sigma} can be used to derive any other case by incorporating the quantum numbers of interest and summing over the  generalized IOS cross-sections
$ \sigma(0  \to K, E)$,
conveniently accessible online. 
Once cross-sections are obtained one should perform an average over the energies  to obtain the rates (see Equation~(\ref{gammak})).    
This enables non-specialized readers to obtain cross-sections/collisional rates for any C$_2$ rotational level within the electronic states  $X \; ^1\Sigma^+_g$   and   $a \; ^3\Pi_u$.

\subsection{Results for $X \; ^1\Sigma^+_g$-state}
\begin{figure}[h]
\begin{center}
 \includegraphics[width=8.5 cm]{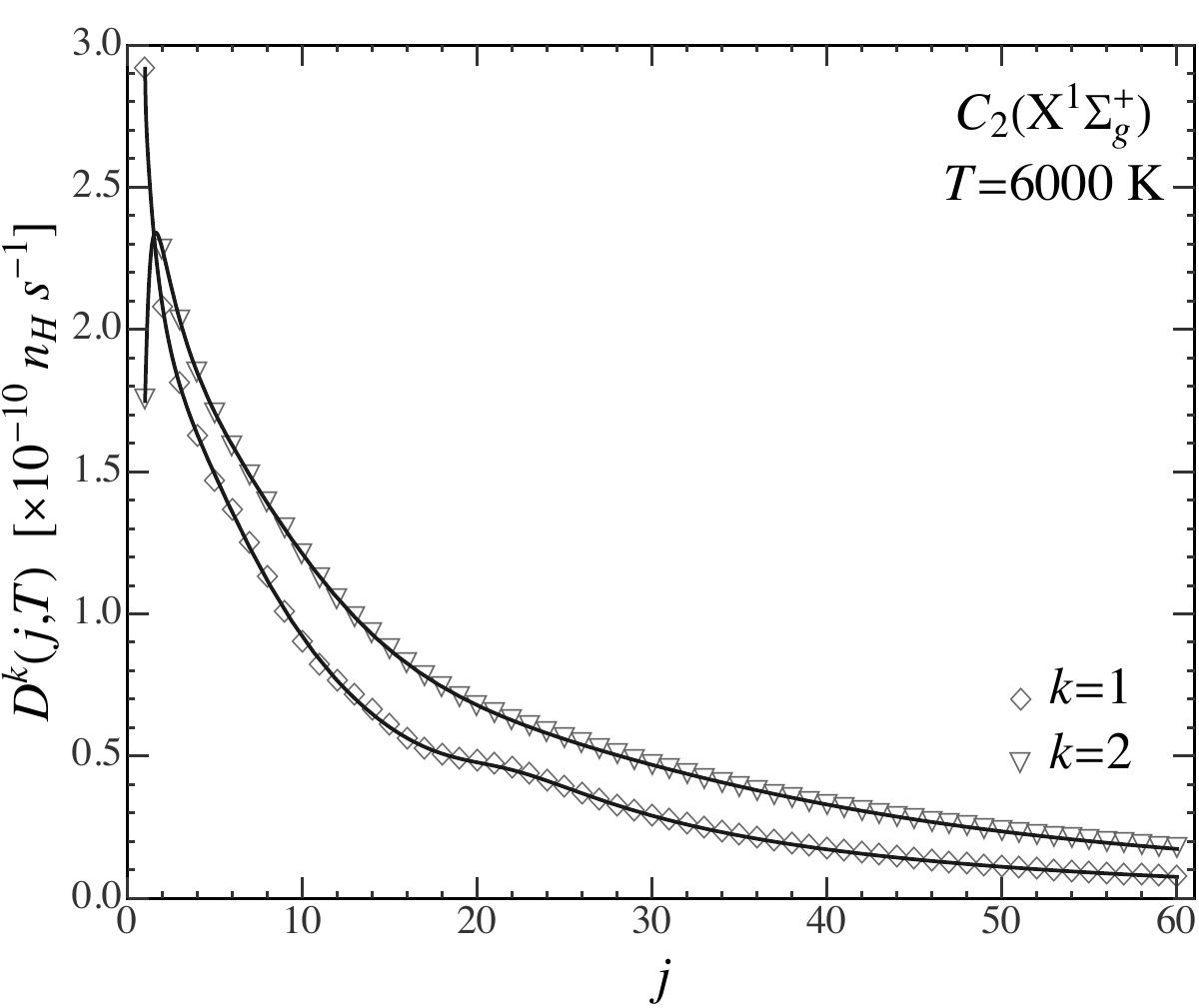}  
  \includegraphics[width=8.5 cm]{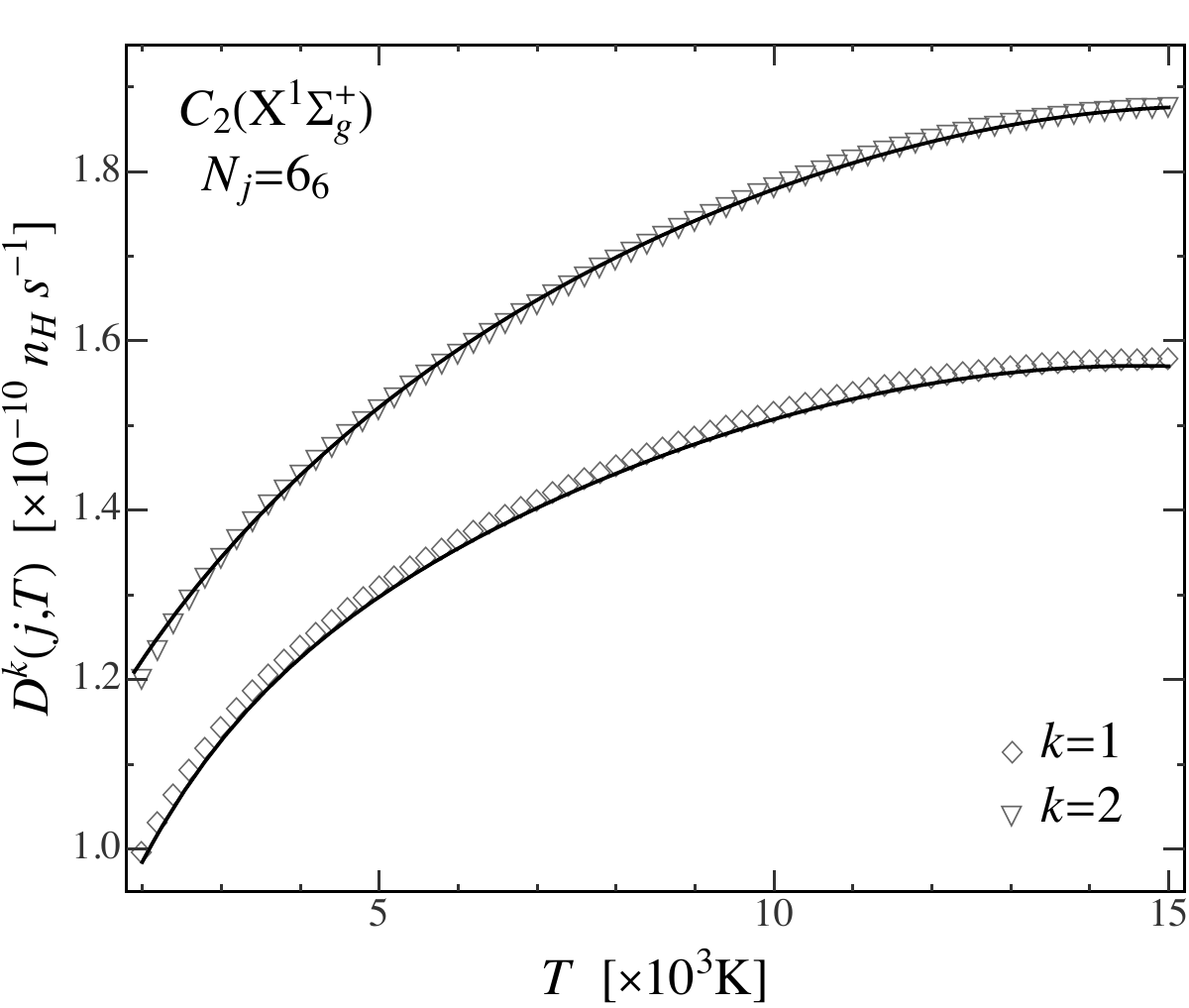}  
 \caption{Collisional depolarization rates, $D^k(j, T) $, for C$_2$ rotational levels within the electronic states  $X \; ^1\Sigma^+_g$. The variation of the rates with $j$ (upper panel) and with $T$ (lower panel) are shown for $k \!=\! 1$ (open diamonds) and $k \!=\! 2$ (open triangles). The 
solid curves show the GP fit values obtained using Equation~(\ref{eq_GP}) and Table~\ref{tab:sigma_depolarization}.}
\label{Dk_j_T6000_sigma}  
\end{center}
\end{figure}

Figure~\ref{Dk_j_T6000_sigma} shows the variation
of
the collisional depolarization rates $D^k(j, T)$ for the alignment ($k\!=\!2$)  and orientation ($k\!=\!1$)   as  functions of $j$ at $T$ = 6,000~K in the upper panel 
and
as  functions of $T$ for the level $N_j$ = $6_6$ in the lower panel.
The depolarization rates 
with tensorial order $k \!=\! 2$ are larger than those with tensorial order $k \!=\! 1$, as seen in Figure~\ref{Dk_j_T6000_sigma}.  As one would expect, 
the $D^k(j, T)$  rates increase with temperature (for a given $j$) 
and 
decrease with increasing $j$ (for a given $T$)   (see e.g. Derouich 2006).
The  $D_{GP}^k(j, T)$ rates
determined by Equation~(\ref{eq_GP}) and Table~\ref{tab:sigma_depolarization} are shown by the 
solid curves in Figure~\ref{Dk_j_T6000_sigma},
and they show excellent agreement with the $D^k(j, T)$  rates that were computed directly. 
The  percentage error  on the $D^k_{GP}(j, T)$ values  is less than 5\% for any $j$ and $T$ in the considered ranges.

\begin{figure}[h]
\begin{center}
  \includegraphics[width=8.5 cm]{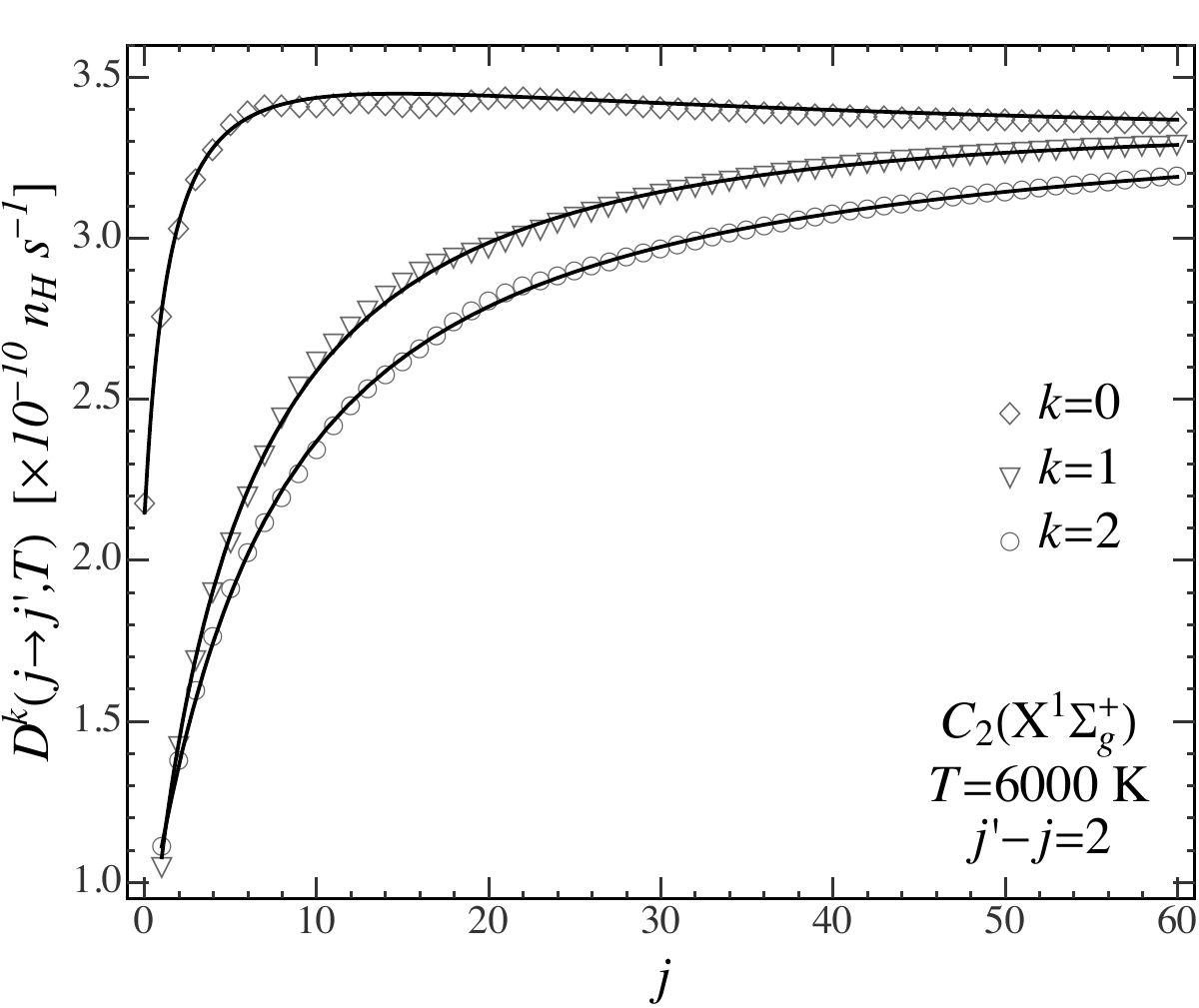}  
 \includegraphics[width=8.5 cm]{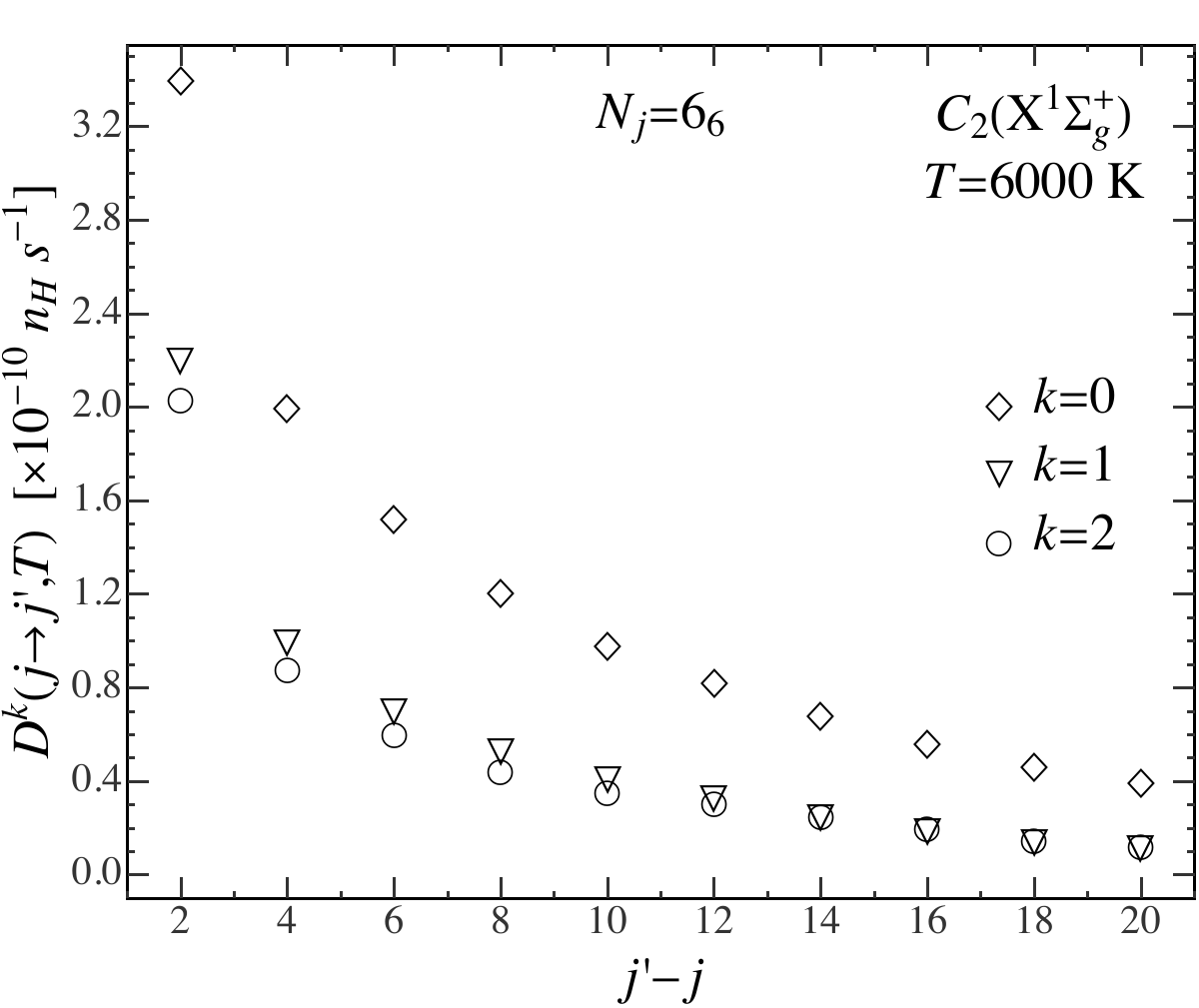}  
 \caption{Collisional transfer rates, $D^k(j  \to  j', T) $, for C$_2$ rotational levels within the electronic states  $X \; ^1\Sigma^+_g$. The variation of the rates  for $k \!=\!0$ (open diamonds), $k \!=\!1$ (open triangles), and  $k \!=\!2$ (open circles) are shown in the upper panel as functions of $j$ for    $ j'\!-\!j \!=\! 2$ and $T$ = 6,000~K and as functions of $ j'\!-\!j$ for the level $N_j$ = $6_6$ and $T$ = 6,000~K in the lower panel. The  solid curves in the upper panel show the GP fit values obtained using Equation~(\ref{eq_GP}) and Table~\ref{tab:sigma_transfer}.}
\label{Fig-Ck_J}  
\end{center}
\end{figure}

In the upper panel of Figure~\ref{Fig-Ck_J}, 
we set $\Delta j \!=\! j'\!-\!j \!=\! 2$  and $T$ = 6,000~K
and 
we show the dependence of the excitation (i.e. $E_{j'} \!>\! E_{j}$)
transfer of polarization rates $D^k(j  \to  j', T) $ for $k \!=\! 0$,  $k \!=\! 1$, and $k \!=\! 2$  
as functions of $j$.
The  solid curves in the  upper panel of Figure~\ref{Fig-Ck_J} show the excellent agreement  between the real values calculated directly and the GP fit values obtained using Equation~(\ref{eq_GP}) and Table~\ref{tab:sigma_transfer}.   
The percentage 
of difference between the real calculated values and the GP values is less than 5\%. 
As it is shown 
in the upper panel of Figure~\ref{Fig-Ck_J},  
the $D^k(j \to j'=j+2, T)$ rapidly increase
with $j$ for low values of $j$ and vary slowly for sufficiently large $j$. $D^k(j \to j'=j+2, T)$ can be considered practically constant for sufficiently large $j$ values. 
We show 
in the lower panel of Figure~\ref{Fig-Ck_J} how the rates $D^k(j  \to  j', T) $  decrease quickly  with increasing of $j'\!-\!j$  for the level $N_j \!=\! 6_6$ and $T$ = 6,000~K.

\subsection{Results for $a \; ^3\Pi_u$-state}
\begin{figure}[h]
\begin{center}
 \includegraphics[width=8.5 cm]{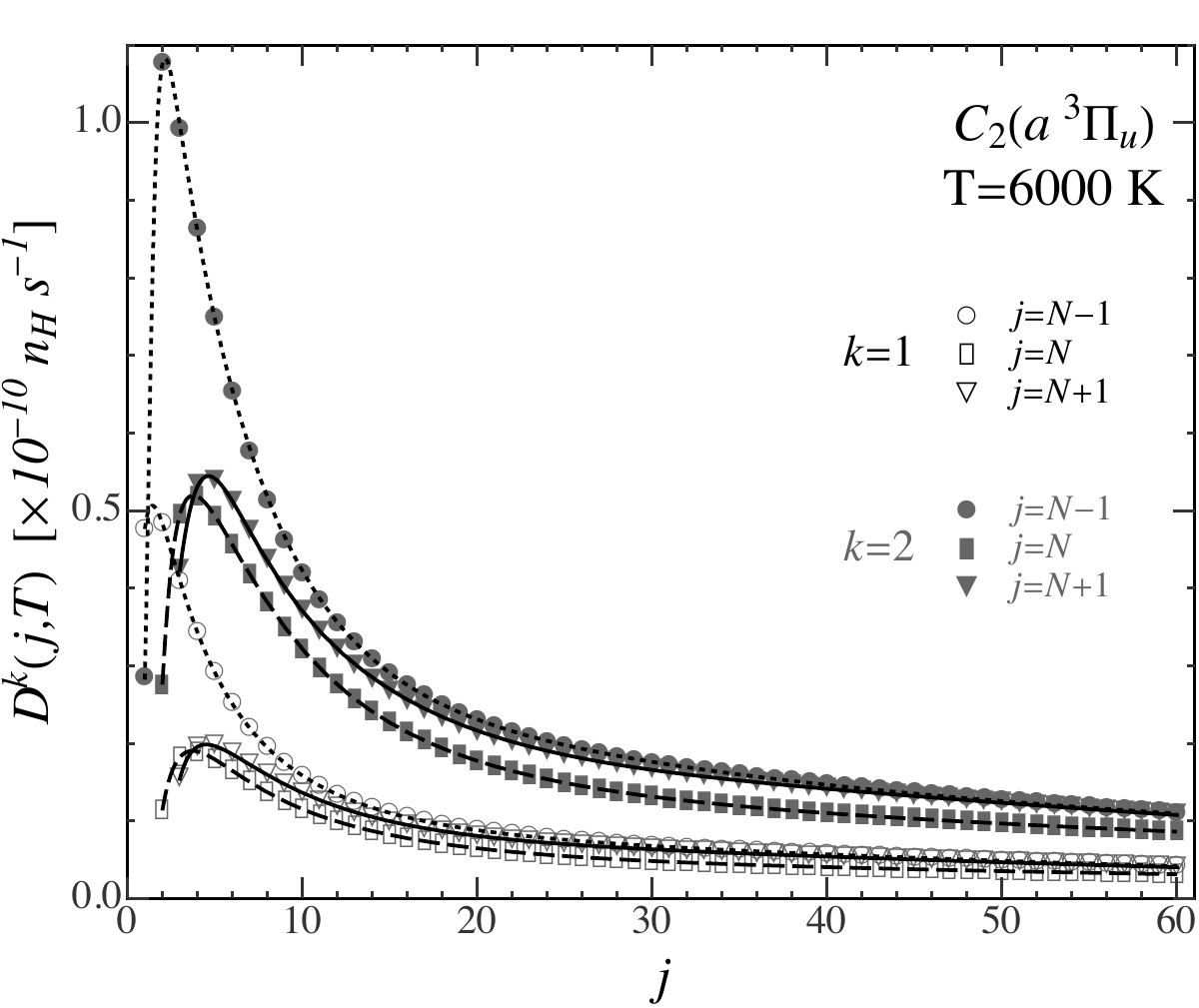}  
  \includegraphics[width=8.5 cm]{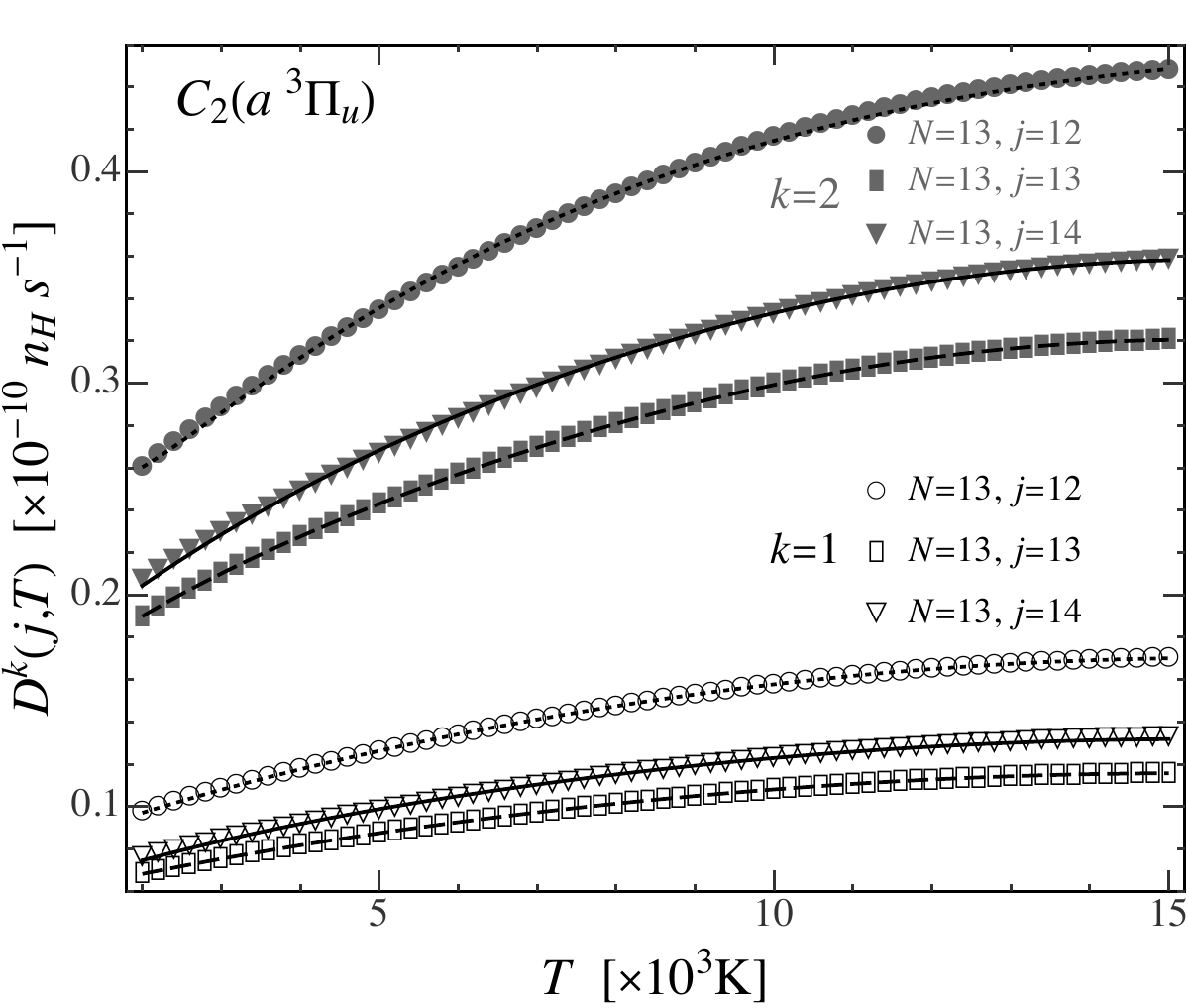}  
 \caption{Collisonal depolarization rates, $D^k(j, T) $, for C$_2$ rotational levels in electronic states $a \; ^3\Pi_u$. The upper panel illustrates the rates for $k \!=\! 1$ (open markers) and $k \!=\! 2$ (solid markers) with respect to $j$ at $T$ = 6,000~K, where $j \!=\! N \!-\! 1$ (circles), $j \!=\! N$ (rectangles), and $j \!=\! N \!+\! 1$ (triangles) are displayed. The lower panel shows the temperature variation of the rates for $k \!=\! 1$ (open markers) and $k \!=\! 2$ (solid markers) with the different $j$ values of the $N \!=\! 13$ multiplet. Both panels display fitted values (dotted, dashed, and solid curves) obtained using Equation~(\ref{eq_GP}) and  GP coefficients of Tables~\ref{tab:Pi_Dk_N_1}, \ref{tab:Pi_Dk_N}, and \ref{tab:pi_j_N+1}.}
\label{Fig-Dk_pi}  
\end{center}
\end{figure}

For the electronic state $a ^3\Pi_u$, with molecular spin $S_d=1$, $j$ can take values of $N \!-\! 1$, $N$, or $N \!+\!1$.
As illustrated in the upper panel of Figure~\ref{Fig-Dk_pi}, for $j \!\gtrsim\! 5$ the depolarization rates, $D^k(j, T) $, decrease with increasing $j$ for constant temperature.
Additionally, the depolarization rates increase as $T$ increases for constant  $j$ as it can be seen in the lower panel of Figure~\ref{Fig-Dk_pi}.  
We note that 
the depolarization rates with tensorial order $k$ = 2 are larger than those with tensorial order $k$ = 1. 
This is also the case for the state  $X \; ^1\Sigma^+_g$.
By using GP fitting techniques, analytical expressions for  $D^k(j, T) $  rates were obtained. These are given 
by Equation~(\ref{eq_GP})  in the temperature range 2,000~--15,000~K and for total angular momentum $j$ going from 1 to 60
 with the GP coefficients provided in Tables~\ref{tab:Pi_Dk_N_1}, \ref{tab:Pi_Dk_N} and \ref{tab:pi_j_N+1} 
for $j\!=\!N\!-\!1$, $j\!=\!N$ and $j\!=\!N\!+\!1$, respectively.
The percentage error in the GP rates $D^k_{GP} (j, T)$ are less than 5\%.
\begin{figure}[h]
\begin{center}
 \includegraphics[width=8.5 cm]{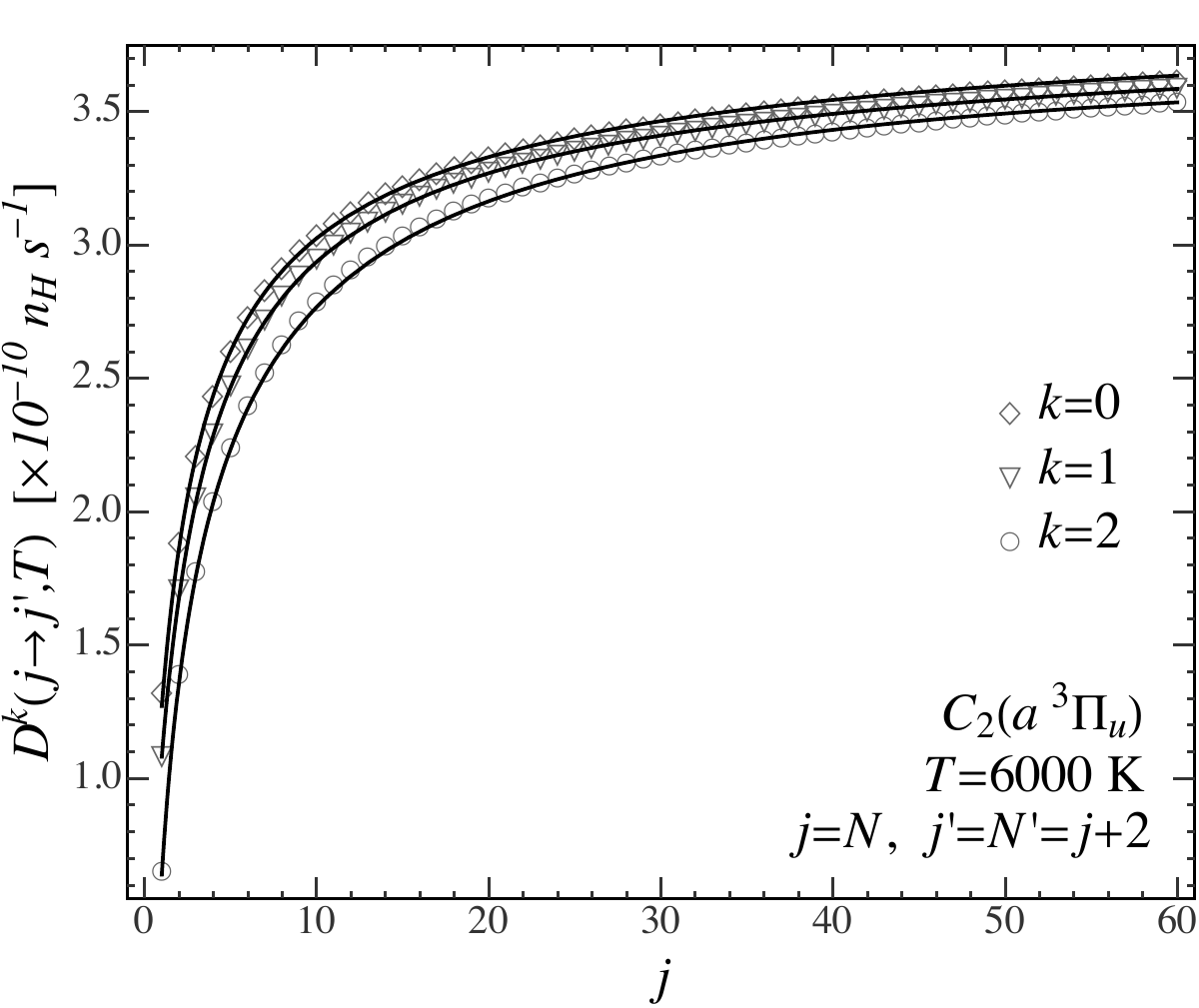}  
  \includegraphics[width=8.5 cm]{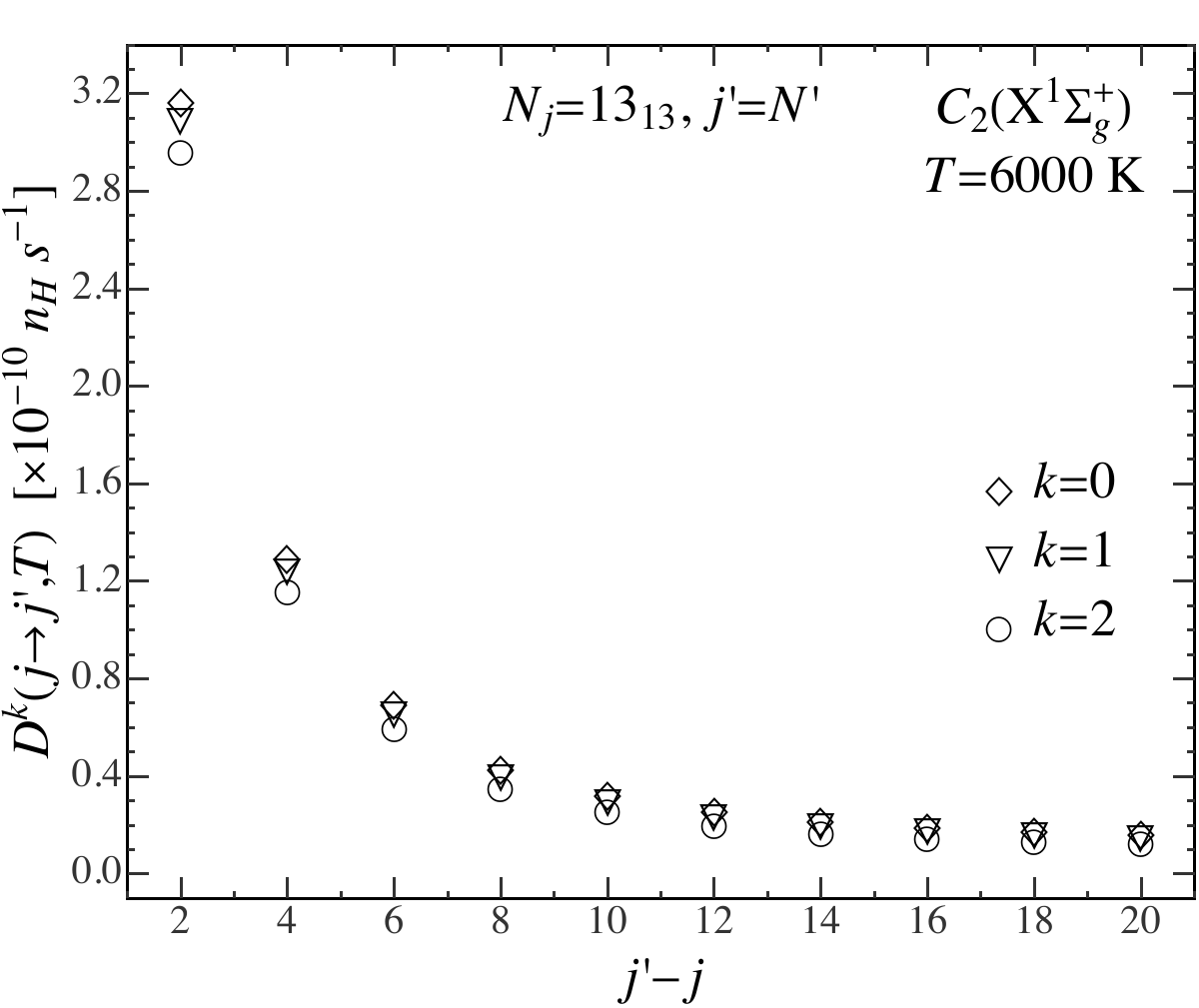}
 \caption{Collisional transfer rates, $D^k(j  \to  j', T) $, for C$_2$ rotational levels within the electronic states  $a \; ^3\Pi_u$. The upper
 panel shows variation with $j$ of the rates for $k \!=\! 0$ (open diamonds), $k \!=\! 1$ (open triangle), and $k \!=\! 2$ (open circles) where we have set   $ j'\!-\!j \!=\! 2$ and $T$ = 6,000~K. The  solid curves show the GP fit values obtained using 
 Equations~(\ref{eq_GP}) and GP coefficients of Table~\ref{tab:Pi_transfer}. The lower panel displays the variation  of the rates  with   $ j'\!-\!j$ for $k \!=\! 0$ (open diamonds), $k \!=\! 1$ (open triangle), and $k \!=\! 2$ (open circles) where we set $T$ = 6,000~K, $N_j$ = $13_{13}$, and $j' \!=\! N'$.  }
\label{Fig-Dk_transfer_pi}  
\end{center}
\end{figure}

Figure~\ref{Fig-Dk_transfer_pi} shows the rates of polarization transfer associated with C$_2$ rotational levels within the electronic states  $a \; ^3\Pi_u$.
The upper panel of Figure~\ref{Fig-Dk_transfer_pi} illustrates significant increase in the transfer rates $D^k(j \to j'\!=\!j\!+\!2, T)$ as $j$ increases for sufficiently small $j$. For $j \gtrsim 20$, the transfer rates keep increasing with increasing $j$ but rather slowly.
This
result gives interesting insights on the differential effect of collisions when comparing the polarization of molecular lines with different $j$-values.
In fact, for lines involving levels with sufficiently large $j$, the $D^k(j \to j'\!=\!j\!+\!2, T)$  can be considered practically constant which greatly simplify the modeling. This is also the case for the electronic state  $X \; ^1\Sigma^+_g$.

\section{Why use  IOS approach in the collision dynamics?}
Unlike close-coupling (CC) methods, which are both time-intensive and require case-by-case calculations, the IOS approach allows for the derivation of comprehensive tables containing generalized IOS cross-sections. These tables facilitate the efficient generation of collisional rates for all transitions and tensorial orders $k$, with good accuracy, particularly at solar temperatures ($T>$5000 K).

\begin{figure}[h]
\begin{center}
\includegraphics[width=8.5 cm]{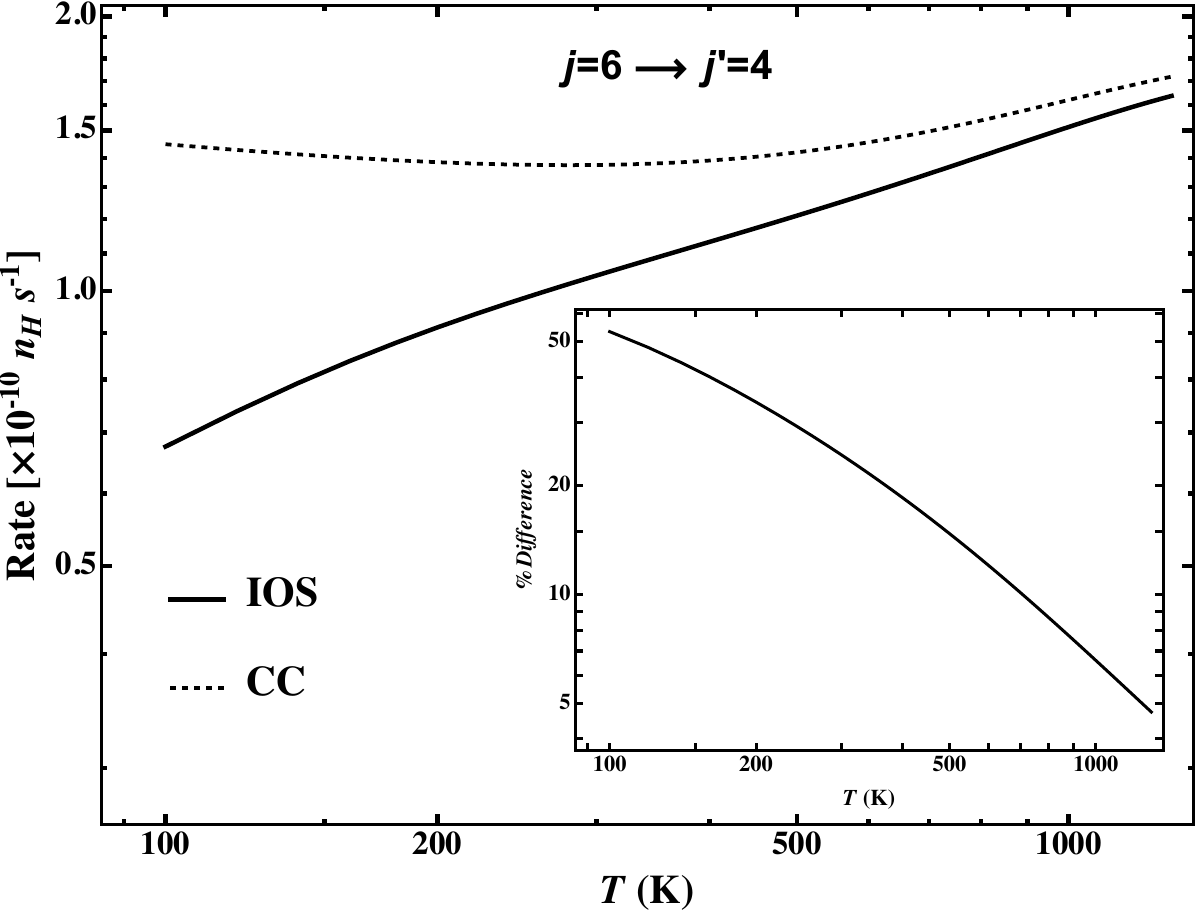}  
\caption{Comparison between the IOS rate (solid curve) and the CC rate (dashed curve) for the collisional rotational de-excitation, $j\!=\!6 \!\to\! j'\!=\!4$ as functions of temperature. In addition, an inset figure is provided to focus on the  \% difference  between the two rates.  The \% difference $  \vert\frac{CCrate-IOSrate}{CCrate}\vert \times 100$ drops to less than $5\%$ at $T\!=\!1300~K$.}
\label{Rates_CC_vs_IOS}  
\end{center}
\end{figure}

To assess the validity of the IOS approximation, we carry out a CC calculation of cross sections using the PES of the interaction C$_2$($X ^1\Sigma^+_g$)+H$^2S$ for kinetic energies upto 5000 $cm^{-1}$ which allows the calculation of collisional rates upto $1300 \, K$ with accuracy better than few percent. In Figure~\ref{Rates_CC_vs_IOS}, we compare the IOS rate to CC rate for the collisional rotation transition $j\!=\!6 \!\to\! j'\!=\!4$. The IOS collisional rate, which is smaller for low temperatures, converges to the CC collisional rate as temperature increases; the difference between the two rates becomes less than $5\%$ for $T=1300~K$ (see Figure~\ref{Rates_CC_vs_IOS}). For higher temperatures, in particular solar temperature, the difference should be negligibly small. This should also to hold true for the PESs arising from interaction between C$_2$($a ^3\Pi_u$) and H ($^2S$).

We also compare our IOS collisional rotational de-excitation cross sections, obtained with our PES for C$_2$ ($X ^1\Sigma^+_g$) + H($^2S$), to the corresponding coupled-state cross sections of Najar et al. (2014), calculated using their own PES.  In particular,  at an energy of approximately 350 cm$^{-1}$, we find:
\begin{itemize}
    \item For the transition $j = 2 \to j' = 0$, our cross section is $1.53~\mathrm{\AA}^2$, while the cross section of Najar \textit{et al.} (2014) is $2~\mathrm{\AA}^2$.
    \item For $j = 4 \to j' = 0$, our cross section is $0.75~\mathrm{\AA}^2$, while the cross section of Najar \textit{et al.} (2014) is $0.8~\mathrm{\AA}^2$.
\end{itemize} 
The differences are smaller than 25\% at these relatively low energies and are expected to become negligible for the higher energies that contribute to the rates at temperatures above 2000~K, which are the focus of this paper. It should be noted that Najar \textit{et al.} (2014) did not consider energies above 350~cm$^{-1}$, and therefore, a comparison at higher energies is not possible.

\section{Solar implications}
Observations  of the second solar spectrum (SSS) revealed the existence of prominent linear polarization signals due to lines of the C$_2$ molecule (e.g. Gandorfer 2000; Faurobert    \& Arnaud  2003; Gandorfer et al. 2004; Kleint et al. 2008). Further, theoretical analyses pointed out the  suitability of these lines for the application of the differential Hanle effect to study  variations of turbulent magnetic fields in the 
photosphere where C$_2$   lines region around 5141 \AA\  form  
(e.g. Berdyugina \& Fluri 2004; Kleint et al. 2010; Kleint et al.  2011;  Mili\'c  \& Faurobert 2012). 
Nevertheless, such
theoretical studies faced the problem of the total lack of collisional rates which impacted the accuracy of their conclusions.

We consider the implication of our results for the molecular C$_2$ lines of the Swan system  ($d ^3\Pi_u$ -- $a ^3\Pi_u$). 
In particular, 
we select the triplet $\text{R}_1(14)$, $\text{R}_2(13)$ and $\text{R}_3(12)$ of the R-branch and
the P-triplet $\text{P}_1(42)$, $\text{P}_2(41)$, $\text{P}_3(40)$ which are suitable for solar magnetic field diagnostics 
(see Kleint et al. 2010).  
To estimate the effect of isotropic collisions, 
we compare the collisional depolarization rates $D^2(j, T)$ of the lower state $a ^3\Pi_u$  
for typical photospheric hydrogen density 
(n$_H$~=~10$^{15}$~--~10$^{16}$~cm$^{-3}$),
to the inverse lifetime (1/$t_{\textrm {life}}$ = $B_{\ell u} I(\lambda_{\ell u})$) of the lower levels of the R-triplet  and P-triplet  lines. 
Here
$I(\lambda_{\ell u})$ denotes the intensity of light of wavelength $\lambda_{\ell u}$ at 
the center of
the solar disk incident on the C$_2$ molecules,
and 
$B_{\ell u} = (w_u/w_{\ell}) (c^2/2h\nu_{u \ell}^3) A_{u \ell}$ denotes the Einstein coefficient for absorption with  $A_{u \ell}$ being the transition probability per unit time for spontaneous emission, $\nu_{u \ell}$  the line frequency, $w_u$ and $w_{\ell}$ the statistical weights of upper and lower levels, $h$ the Planck's constant, and $c$ the speed of light.
We note that 
the rate of radiative relaxation from the electronic state $a ^3\Pi_u$ to the lower energy electronic state $X \; ^1\Sigma^+_g$ is negligibly small (see e.g. Wehres et al. 2010).

In Table~\ref{tab:sol_impli}, we show $B_{\ell u} I(\lambda)$ and the linear depolarization rates, $D^{k=2}(j_{\ell})$, calculated at the effective photospheric temperature, $T_{\rm eff} \!=\! 5778$~K, and at typical values of Hydrogen density $n_{H} \!=\! 10^{15} {\rm cm}^{\text{-}3}$
and
$n_{H} \!=\! 10^{16} {\rm cm}^{\text{-}3}$ 
in the photosphere.
The values of the core relative intensity of the absorption lines are taken from the solar atlas of Delbouille et al. (1972) whereas the corresponding absolute continuum values are interpolated from the data given in  Allen  (1976). The values of the Einstein $A_{u \ell}$ coefficients   are derived from  Kleint et al. (2010).

In the case of the  R-triplet, 
the linear depolarization rates  $D^{k=2}$ are 
roughly 
$\frac{1}{3} B_{\ell u} I(\lambda_{\ell u})$ for $n_{H} \!=\! 10^{15}$~cm$^{\text{-}3}$ which means that the lower levels of the R-triplet lines residing within the electronic state C$_2$~$a ^3\Pi_u$ should be affected by the depolarizing collisions. 
On the other hand, 
for $n_{H} \!=\! 10^{16}$~cm$^{\text{-}3}$, 
the linear depolarization rates $D^{k=2}$ of the lower levels for lines of the R-triplet are 
roughly 
$3 \ B_{\ell u} I(\lambda_{\ell u})$ 
which renders the depolarizing effect of collisions stronger. 
Nevertheless, 
for both perturbers' densities the depolarizing collisional rates are not sufficiently high to completely depolarize the lower levels of the R-triplet lines.

Similarly for the P-triplet case in the typical photospheric conditions, 
the lower levels for lines of the triplet cannot be completely depolarized by collisions given that
the collisional depolarization rates $D^{k=2}$ of these levels,
which are relatively lower given the relatively larger $j$ values (see the upper panel of Figure~\ref{Fig-Dk_pi}  
),
 are comparable to their inverse lifetime, $B_{\ell u} I(\lambda_{\ell u})$, (see Table~\ref{tab:sol_impli}).

It is clear that with the typical photospheric densities,
$n_{H} \!=\! 10^{15}$--$10^{16} \; {\rm cm}^{\text{-}3}$, 
collisions with hydrogen atoms partially depolarize the rotational levels 
of 
the lower electronic level of the  C$_2$ lines of the   Swan
system  ($d ^3\Pi_u$ -- $a ^3\Pi_u$). 
Hence,
one has to incorporate the collisional depolarization rates when solving the  SEE for the polarization of observed lines. 

The R-branch lines, $\text{R}_1(14)$, $\text{R}_2(13)$, and $\text{R}_3(12)$, are more significantly affected by collisions compared to the P-branch lines, $\text{P}_1(42)$, $\text{P}_2(41)$, and $\text{P}_3(40)$, because the latter have larger $j$-values. As demonstrated in the previous section, the collisional effect decreases as $j$ increases.
\begin{table*}
\centering
\caption{Comparison between the linear depolarization rates $D^2$ of the C$_2$~$a ^3\Pi_u$ state to its inverse lifetime  ${1}/{t_{life}}$=$B_{\ell u} I(\lambda)$.  Note that $I(\lambda_{u l}) $ is given in $(10^{-5} {\rm erg} $  ${\rm cm}^{-2} \,{\rm s}^{-1} {\rm sr}^{-1} {\rm Hz}^{-1})$. The Lines are, respectively, $\text{R}_1(14)$, $\text{R}_2(13)$, $\text{R}_3(12)$, $\text{P}_1(42)$, $\text{P}_2(41)$ and $\text{P}_3(40)$.}
\begin{tabular}{c c c c c c c c  } 
\hline
\hline
  $\lambda_{u l} \, (\AA)$ & $j $ & $I(\lambda_{u l}) \,   $ &   $B_{l u} I(\lambda_{u l})$ &
\multicolumn{2}{c}{$D^{2}( j,T \!=\! 5778~\rm{K}) \ (10^{5} {\rm s}^{-1})$}
\\
    &   &      &    $(10^{5} {\rm s}^{-1})$ &   $n_{\rm H} \!=\! 10^{15} {\rm cm}^{-3}$  &  $n_{\rm H} \!=\! 10^{16} {\rm cm}^{-3}$
\\
\hline
  $5139.93$ & 14  &   $8.13008$ & $0.845676$ &   $0.279154$ & $2.79154$ \\
  $5140.14$ & 13  &   $8.62069$ & $0.917542$ &   $0.253947$ & $2.53947$ \\
  $5140.38$ & 12  &   $9.34579$ & $1.014485$ &   $0.350343$ & $3.50343$ \\
  $5141.21$ & 42  &   $8.26446$ & $0.736174$ &   $0.136336$ & $1.36336$ \\
  $5141.19$ & 41  &   $8.47458$ & $0.754441$ &   $0.107232$ & $1.07232$ \\
  $5141.31$ & 40  &   $8.69565$ & $0.834171$ &   $0.145746$ & $1.45746$ \\       
\hline
\hline
\end{tabular}
\label{tab:sol_impli}
\end{table*}
%
%

\section{Conclusion}
This paper continues a series of investigations concerned with the collisional depolarization of spectral lines of solar molecules such as MgH, CN, and C$_2$. 
In this study, 
we have computed 
the quantum collisional depolarization and polarization transfer rates for
C$_2$ $(X ^1\Sigma^+g, a ^3\Pi_u)$ + H$(^2S{1/2})$  
isotropic collisions. 
The computation involved calculating the potential energy surfaces (PESs) using the MOLPRO package, followed by solving the quantum dynamics using the MOLSCAT code. 
Sophisticated genetic programming techniques 
were 
employed to derive analytical expressions for the temperature and total molecular angular momentum dependencies
of 
the collisional depolarization and polarization transfer rates. 
The results show that isotropic collisions with neutral hydrogen partially depolarize the lower state of C$_2$ lines, implying the limitations of neglecting lower-level polarization. 
Collisional depolarization  and polarization transfer
rates are a fundamental ingredient for interpreting C$_2$ polarization in terms of magnetic fields in the quiet regions of the Sun.

\begin{acknowledgements}  This research work was funded by Institutional Fund Projects under grant no. (IFPIP:772-130-1443). The authors gratefully acknowledge technical and financial support provided by the Ministry of Education and King Abdulaziz University, DSR, Jeddah, Saudi Arabia. We thank François Lique for his insightful discussion  on molecular collision physics. 
\end{acknowledgements}
%
%


\begin{appendix}
\section{Different Tables giving the GP coefficients}
\begin{table*}[h]
\centering
\caption{GP coefficients corresponding to  the analytical expression of Equation~(\ref{eq_GP}) for  $D^k(j \!=\! N, T)$. Coefficients are given for both $k$ = 1 and $k$ = 2 for rotational levels of the  $\Sigma$-state.
 }
\begin{tabular}{c c c c c c c } 
\hline
\hline
  &  $a_i^k$   &  $\alpha_i^k$  &  $\beta_i^k$ &  $b_i^k$  &   $\gamma_i^k$  &  $\delta_i^k$  \\
\hline
$i$  & & & {\centering $k=1$}\\
\hline
$1$ & $\,1.991191e{\text{-}3}$  &  -$0.581446$    & -$0.3057279$    &  $\,2.775182e{\text{-}8}$  & $4.8698780$  &  -$0.9948641$  \\
$2$ & $\,0.9256007$             &  $\,1.563707$   & -$0.1573567$    &  $\,3.318555e{3}$          & $0.7634837$  &  -$0.7735423$  \\
$3$ & -$0.9256059$              &  $\,1.563705$   & -$0.1573564$    & -$3.823419e{3}$            & $0.7683499$  &  -$0.7727445$  \\
$4$ & -$1.272366e{\text{-}4}$   &  $\,1.524807$   &  $\,0.9668442$  &  $\,6.282801e{2}$          & $0.8160999$  &  -$0.7646596$  \\
$5$ & $\,2.845039e{\text{-}4}$  &  $\,1.525127$   &  $\,0.9668662$  & -$1.782060e{2}$            & $0.9014560$  &  -$0.7490103$  \\
$6$ & -$1.572672e{\text{-}4}$   &  $\,1.525386$   &  $\,0.9668841$  &  $\,5.486061e1$            & $0.9463913$  &  -$0.7401261$  \\
$7$ & $\,8.843118e{\text{-}9}$  &  -$35.64545$    &  $\,1.0808530$  & $\,3.048508e{\text{-}7}$   & -$1.8538130$ &  $\,0.9168674$ \\
$8$ & $\,2.461312e{\text{-}8}$  &  -$2.357598$    &  $\,1.3237760$  & $\,1.441014e{\text{-}21}$  & -$1.7989560$ &  $\,4.1172120$ \\
\hline
$i$  & & & $k=2$\\
\hline
$1$ &  $\,6.001485e{\text{-}4}$  &    $\,2.396011$  &    $0.5982441$  &    -$2.895777e{\text{-}3}$    &    $\,1.752254$  &    $0.8141479$ \\
$2$ &  -$1.117526e{\text{-}3}$   &    $\,2.396003$  &    $0.5984454$  &    $\,2.895777e{\text{-}3}$   &    $\,1.752254$  &    $0.8141479$ \\
$3$ &  $\,5.173784e{\text{-}4}$  &    $\,2.395994$  &    $0.5986782$  &    $\,2.718900e{\text{-}14}$  &    $\,4.056257$  &    $0.9387745$  \\
$4$ &  $\,3.39671e{\text{-}10}$  &    -$0.5960658$  &    $1.464536$   &    $\,1.576906e{\text{-}11}$  &    -$1.520357$   &    $2.107281$  \\
$5$ &  $\,1.659791e{\text{-}9}$  &    -$1.824684$   &    $2.816542$   &    $\,5.131261e{\text{-}12}$  &    -$5.888703$   &    $2.205830$  \\
$6$ &  -$1.659459e{\text{-}9}$   &    -$1.824676$   &    $2.816558$   &    $\,7.264044e{\text{-}35}$  &    -$2.347249$   &    $7.450174$   \\
\hline
\hline
\end{tabular}
\begin{flushleft}
\end{flushleft}
\label{tab:sigma_depolarization}
\end{table*}
\begin{table*}[h]
\centering
\caption{
The GP coefficients, as per the analytical expression in Equation (10) for $D^k(j \!=\! N \!\to\!  j' \!=\! N \!+\! 2, T)$, are provided for rotational levels of the $\Sigma$-state, where $k$ =0,  1 and $k$ = 2.
}
\begin{tabular}{c c c c c c c } 
\hline
\hline
  &  $a_i^k$   &  $j^{\alpha_i^k}$  &  $\beta_i^k$ &  $b_i^k$  &   $\gamma_i^k$  &  $\delta_i^k$  \\
\hline
$i$  & & & {\centering $k=0$}\\
\hline
$1$ & $\,1.397544e3$           &  $1.184034$  & -$1.020533$   &  $0.5188084$		         & $1.344210$  &  $2.070097e{\text{-}2}$\\
$2$ & $\,7.100799e{\text{-}2}$ &  $1.598421$  & $\,0.2540003$ &  $0.7106673$             & $1.162418$  &  $0.1598386$\\
$3$ & $\,3.735336e{\text{-}2}$ &  $0.9630494$ & $\,0.7948434$ &  $4.799789e{\text{-}2}$  & $0.8078731$ &  $0.5528220$\\
$4$ & -$4.478655e{\text{-}7}$  &  $1.283280$  & $\,1.599432$  &  $1.244310e{\text{-}6}$  & $0.8860535$ &  $1.561604$\\
$5$ & $\,5.633616e{\text{-}2}$ &  $   0    $  & $\,0.7269272$ &  $0.4886689$             & $    0   $  &  $0.3906900$\\
\hline
$i$  & & & $k=1$\\
\hline
$1$ & -$2.296152$              & $0.7256945$ & -$2.405169e{\text{-}2}$ & $\,0.1039076$            & $\,1.662413$    &  $0.3857427$\\
$2$ & $\,3.809447e{\text{-}2}$ & $1.475278$  & $\,0.8186476$           & $\,1.682204e{\text{-}4}$ & $\,0.8653501$   &  $0.7800596$\\
$3$ & -$1.193235e{\text{-}2}$  & $1.261421$  & $\,1.016065$            & -$1.223617e{\text{-}6}$  & $\,1.379011$    &  $1.540615$\\
$4$ & $\,2.272656e{\text{-}4}$ & $0.3249925$ & $\,1.258974$            & $\,1.693369e{\text{-}5}$ & $\,0.6833434$   &  $1.551339$\\
$5$ & $\,1.588990e{\text{-}5}$ & $0.4005488$ & $\,1.556258$            & $\,2.137974e{\text{-}5}$ & -$5.004082e{\text{-}2}$ & $1.599273$\\
$6$ & $\,3.827193e{\text{-}5}$ &  $1.063432$  & $\,1.629335$            & $\,6.390334e{\text{-}9}$  &  $1.267591$    &  $2.218129$\\
\hline
$i$  & & & $k=2$\\
\hline
$1$ & -$2.145054e{\text{-}11}$ &  $1.109234$  &  $2.934953$  &  $1.143386e{\text{-}3}$  &  $1.156976$  &  $0.9945199$\\
$2$ & $\,1.144444e{\text{-}2}$ &  $0.9603254$ &  $0.9600221$ &  $9.317665e{\text{-}4}$  &  $1.333167$  &  $0.7144136$\\
$3$ & $\,2.813450e{\text{-}4}$ &  $1.228673$  &  $1.047091$  &  $7.713058e{\text{-}3}$  &  $1.062634$  &  $0.7403552$\\
$4$ & $\,1.605891e{\text{-}5}$ &  $1.219828$  &  $1.600001$  &  $3.864867e{\text{-}3}$  &  $0.9338449$ &  $0.8553057$\\
$5$ & $\,5.890345e{\text{-}2}$ &  $    0   $  &  $0.8765076$ &  $1.064727$              &  $   0   $   &  $0.5685439$\\
\hline
\hline
\end{tabular}
\begin{flushleft}
\end{flushleft}
\label{tab:sigma_transfer}
\end{table*}

\begin{table*}[h]
\centering
\caption{
 GP coefficients, corresponding to the  Equation (10) for  $D^k(j \!=\! N \!-\! 1, T)$, are provided for rotational levels within the $\Pi$-state, covering both $k$ = 1 and $k$ = 2.
}
\begin{tabular}{c c c c c c c } 
\hline
\hline
  &  $a_i^k$   &  $j^{\alpha_i^k}$  &  $\beta_i^k$ &  $b_i^k$  &   $\gamma_i^k$  &  $\delta_i^k$  \\
\hline
$i$  & & & {\centering $k=1$}\\
\hline
$1$ & $\,9.631358e{\text{-}4}$  &  $1.270850$  &  $0.9713363$ &  $\,3.821693e{\text{-}7}$  &  $0.477288$   &  $0.3895764$  \\
$2$ & -$1.529505e{\text{-}3}$   &  $1.271256$  &  $0.9713726$ &  $\,2.376594e{\text{-}12}$ &  $3.874105$   &  $0.6974655$  \\
$3$ & $\,5.663722e{\text{-}4}$  &  $1.271943$  &  $0.9714340$ &  $\,1.143000e{\text{-}5}$  &  $1.646778$   &  $1.556904$   \\
$4$ & -$3.359058e{\text{-}5}$   &  $0.5428953$ &  $1.523613$  &  -$1.142996e{\text{-}5}$   &  $1.646778$   &  $1.556904$   \\
$5$ & $\,3.359058e{\text{-}5}$  &  $0.5428957$ &  $1.523613$  &  $\,1.321659e{\text{-}21}$ &  $1.613181$   &  $3.739739$   \\
\hline
$i$  & & & $k=2$\\
\hline
$1$ &  $\,8.921348e{\text{-}5}$  &  $1.216925$  & $1.193702$  &  -$1.536963e{\text{-}2}$   & $\,1.063446$  &  $0.5335961$ \\
$2$ &  -$2.910309e{\text{-}4}$   &  $1.198176$  & $1.194265$  &  $\,1.532168e{\text{-}2}$  & $\,1.062882$  &  $0.5342671$ \\
$3$ &  $\,6.203630e{\text{-}5}$  &  $1.189519$  & $1.194559$  &  $\,5.326371e{\text{-}6}$  & -$4.183293$   &  $0.9498192$ \\
$4$ &  $\,6.870049e{\text{-}5}$  &  $1.189240$  & $1.194568$  &  $\,2.531897e{\text{-}9}$  & $\,2.977568$  &  $0.9846094$ \\
$5$ &  $\,7.120522e{\text{-}5}$  &  $1.189136$  & $1.194572$  &  $\,5.741291e{\text{-}14}$ & $\,1.042410$  &  $2.549029$  \\
\hline
\hline
\end{tabular}
\begin{flushleft}
\end{flushleft}
\label{tab:Pi_Dk_N_1}
\end{table*}
\begin{table*}[h]
\centering
\caption{
GP coefficients, corresponding to the  Equation (10) for  $D^k(j \!=\! N, T)$,  are presented for rotational levels within the $\Pi$-state, where $k$ = 1 and $k$ = 2.
}
\begin{tabular}{c c c c c c c } 
\hline
\hline
  &  $a_i^k$   &  $j^{\alpha_i^k}$  &  $\beta_i^k$ &  $b_i^k$  &   $\gamma_i^k$  &  $\delta_i^k$  \\
\hline
$i$  & & & {\centering $k=1$}\\
\hline
$1$ &   $\,2.624783e{\text{-}10}$ &  $3.280700$  &  $0.7520319$  &  $\,0.2436235$             &  $1.459148$  &  $0.2180211$ \\
$2$ &   $\,7.484584e{\text{-}4}$  &  $0.3670139$ &  $1.120077$   &  -$0.1199928$              &  $1.473343$  &  $0.2248723$ \\
$3$ &   -$3.449391e{\text{-}3}$   &  $0.3609992$ &  $1.121988$   &  -$0.3094506$              &  $1.473360$  &  $0.2248829$ \\
$4$ &   $\,9.381872e{\text{-}4}$  &  $0.3600937$ &  $1.122277$   &  $\,0.1868954$             &  $1.489204$  &  $0.2327578$ \\
$5$ &   $\,8.944660e{\text{-}4}$  &  $0.3592776$ &  $1.122539$   &  $\,2.722244e{\text{-}10}$ &  $4.498694$  &  $0.5419195$ \\
$6$ &   $\,8.683892e{\text{-}4}$  &  $0.3587206$ &  $1.122718$   &  $\,3.437098e{\text{-}20}$ &  $1.959055$  &  $3.747129$  \\
\hline
$i$  & & & $k=2$\\
\hline
$1$ & $\,1.030832e{\text{-}3}$  &  $2.196770$ &  $0.7149692$ &  $4.802783e{\text{-}3}$  &  $\,0.3880369$ &  $0.2112838$ \\
$2$ & -$5.817601e{\text{-}3}$   &  $2.259210$ &  $0.7201542$ &  $6.225231e{\text{-}4}$  &  $\,1.664431$  &  $0.3283672$ \\
$3$ & $\,4.818115e{\text{-}3}$  &  $2.269287$ &  $0.7209898$ &  $6.642306e{\text{-}2}$  &  -$4.561958$   &  $0.3390357$ \\
$4$ & $\,1.956628e{\text{-}5}$  &  $1.665406$ &  $1.619418$  &  $7.858702e{\text{-}10}$ &  $\,5.262455$  &  $0.4312077$ \\
$5$ & -$5.438509e{\text{-}5}$   &  $1.664943$ &  $1.619572$  &  $3.935674e{\text{-}7}$  &  $\,2.940227$  &  $0.9450288$ \\
$6$ & $\,3.482543e{\text{-}5}$  &  $1.664755$ &  $1.619634$  &  $6.684828e{\text{-}21}$ &  $\,2.846936$  &  $4.072948$  \\
\hline
\hline
\end{tabular}
\begin{flushleft}
\end{flushleft}
\label{tab:Pi_Dk_N}
\end{table*}
\begin{table*}[h]
\centering
\caption{
GP coefficients, corresponding to the  Equation (10) for  $D^k(j \!=\! N \!+\! 1, T)$,
 are provided for rotational levels within the $\Pi$-state, with $k$ equal to 1 and 2.}
\begin{tabular}{c c c c c c c } 
\hline
\hline
  &  $a_i^k$   &  $j^{\alpha_i^k}$  &  $\beta_i^k$ &  $b_i^k$  &   $\gamma_i^k$  &  $\delta_i^k$  \\
\hline
$i$  & & & {\centering $k=1$}\\
\hline
$1$ &  $\,2.881511e{\text{-}5}$  &  $0.6858308$  &  $1.280080$  &  -$7.963110e{\text{-}4}$   & $1.455747$ &  $0.8831807$\\
$2$ &  -$9.313931e{\text{-}5}$   &  $0.6702042$  &  $1.280779$  &  $\,7.962892e{\text{-}4}$  & $1.455742$ &  $0.8831870$\\
$3$ &  $\,1.573687e{\text{-}4}$  &  $0.6529715$  &  $1.281580$  &  $\,5.486403e{\text{-}12}$ & $3.246100$ &  $0.9696506$ \\
$4$ &  -$9.304530e{\text{-}5}$   &  $0.6458879$  &  $1.281917$  &  $\,9.461886e{\text{-}25}$ & $1.289088$ &  $4.593579$\\
\hline
$i$  & & & $k=2$\\
\hline
$1$ & $\,7.778896e{\text{-}5}$  &  $1.148716$  &  $1.056787$ &  $\,1.469117e{\text{-}7}$ &  $\,2.926629$ &  $0.6638563$ \\
$2$ & -$2.744207e{\text{-}4}$   &  $1.062771$  &  $1.070655$ &  -$1.371769e{\text{-}2}$  &  $\,1.264619$ &  $0.6899927$  \\
$3$ & $\,3.058522e{\text{-}5}$  &  $1.033921$  &  $1.075424$ &  $\,1.636800e{\text{-}2}$ &  $\,1.282435$ &  $0.6957809$  \\
$4$ & $\,1.687107e{\text{-}4}$  &  $1.015060$  &  $1.078548$ &  $\,4.727379e{\text{-}3}$ &  -$2.247660$  &  $0.7048250$  \\
$5$ & -$7.833470e{\text{-}18}$  &  $0.1888026$ &  $3.768634$ &  -$2.842829e{\text{-}3}$  &  $\,1.348965$ &  $0.7114709$  \\
\hline
\hline
\end{tabular}
\begin{flushleft}
\end{flushleft}
\label{tab:pi_j_N+1}
\end{table*}

\begin{table*}[h]
\centering
\caption{
GP coefficients associated to Equation (10) for $D^k(j \!=\! N \!\to\! j' \!=\! N' \!=\! N \!+\! 2, T)$ are given for rotational levels within the $\Pi$-state, where $k$ equals 0, 1 and 2.
}
\begin{tabular}{c c c c c c c } 
\hline
\hline
  &  $a_i^k$   &  $j^{\alpha_i^k}$  &  $\beta_i^k$ &  $b_i^k$  &   $\gamma_i^k$  &  $\delta_i^k$  \\
\hline
$i$  & & & {\centering $k=0$}\\
\hline
$1$ &  $8.131733e{\text{-}2}$ & $1.368889$  &  $0.4501252$ &  $2.579855e{\text{-}2}$  &  $2.216276$  & -$0.2126671$\\
$2$ &  $1.854918e{\text{-}3}$ & $1.481682$  &  $0.6827418$ &  $0.1051687$             &  $1.231380$  &  $\,0.4790863$\\
$3$ &  $7.145952e{\text{-}4}$ & $1.226836$  &  $1.089026$  &  $3.530731e{\text{-}2}$  &  $0.2205783$ &  $\,0.6630505$\\
$4$ &  $6.316946e{\text{-}4}$ & $1.250177$  &  $1.112160$  &  $1.052831e{\text{-}7}$  &  $1.207201$ &   $\,1.834124$ \\
\hline
$i$  & & & $k=1$\\
\hline
$1$ & -$9.236490e{\text{-}11}$ &  $0.9900076$ &  $2.855520$ &  $3.553307e{\text{-}3}$ &  $0.9487627$ &  $0.8797230$\\
$2$ & $\,6.004538e{\text{-}5}$ &  $0.9881903$ & $\,1.302873$ &  $1.111160e{\text{-}2}$ &   $1.078368$ & $0.6830387$\\
$3$ & $\,4.043410e{\text{-}5}$ &  $\,1.088956$ & $1.428987$ &  $1.273399e{\text{-}2}$ &  $0.8321978$ &  $0.6906823$\\
$4$ & $\,4.497109e{\text{-}6}$ &  $0.9870536$ &  $1.759647$ & $5.572296e{\text{-}3}$ &  $0.9256184$ & $0.8196994$ \\
$5$ & -$8.575254e{\text{-}6}$ &     $0$ &  $1.555283$  &  $1.402920e{\text{-}2}$ &  $1.057578$ & $0.6979735$\\
$6$ & $\,1.971299e{\text{-}2}$ &  $\,1.010017$  & $0.9322726$  &  $0.2245883$ &  $0$ &  $0.6259280$\\
\hline
$i$  & & & $k=2$\\
\hline
$1$ & $\,4.613728e{\text{-}2}$ & $1.007588$ & $0.8005586$ & $0.2866146$ &  $1.144674$ &  $0.1364757$\\
$2$ & -$2.640124e{\text{-}2}$ &  $0.8563931$ &  $0.8711782$ &  $4.477300e{\text{-}2}$ & 
   $0.4118615$ &  $0.4724249$\\
$3$ & $\,8.9231410e{\text{-}6}$ &  $1.019305$ &  $1.596336$  & $8.635446e{\text{-}3}$ &  $1.130790$ &  $0.6933346$\\
$4$ & -$4.865965e{\text{-}11}$ &  $1.043426$ & 
 $2.764539$ &  $1.102920e{\text{-}3}$ &  $0.7979468$ & $0.8035571$\\
\hline
\hline
\end{tabular}
\begin{flushleft}
\end{flushleft}
\label{tab:Pi_transfer}
\end{table*}
\end{appendix}
\end{document}